\documentclass[bibyear]{aa} 

\usepackage{natbib}
\setcitestyle{round,aysep={},yysep={;}}
\usepackage{graphicx}
\usepackage{txfonts}
\begin{document}

   \title{Distribution of red clump stars does not support the X-shaped Galactic bulge}
   \subtitle{}

   \author{M. L\'opez-Corredoira\inst{1,2}, 
Y.-W. Lee\inst{3}, F. Garz\'on\inst{1,2}, D. Lim\inst{3}}

   \institute{$^1$ Instituto de Astrof\'\i sica de Canarias, 
E-38205 La Laguna, Tenerife, Spain\\
$^2$ Departamento de Astrof\'\i sica, Universidad de La Laguna,
E-38206 La Laguna, Tenerife, Spain \\
$^3$ Center for Galaxy Evolution Research \& Department of Astronomy, 
Yonsei University, Seoul 03722, Republic of Korea}

   \date{Received xxxx; accepted xxxx}

 
  \abstract
  {Claims of an X-shaped Galactic bulge were based on the assumption of red clump stars as standard candles
in some lines of sight crossing the off-plane bulge. However, some doubts have been cast on whether the two peaks in star counts along the line of sight really represent a double peak in the density distribution, or whether there is something wrong with the assumption of a unique constant absolute magnitude for all of these stars.}
   {With the advent of Gaia-DR2 parallaxes in combination with near-infrared VISTA-VVV data, 
we are able to check which of the hypotheses is correct.}
   {We calculated the median absolute magnitude $M_K$ corresponding to both peaks of putative red clumps in seven lines of sight with the lowest extinction in the interesting coordinates' range.}
   {The difference between the absolute magnitude of the bright and the faint peak is $\Delta M_K\approx 0.4$. The selected stars in both peaks cannot be represented by the same red clump giants with constant $M_K\approx -1.6$.}
   {The hypothesis that the bulge contains an X-shape is based on the assumption that the faint and bright peaks of the density distribution towards the bulge are dominated by standard red clump stars. However, we show that both the faint and bright peaks cannot be dominated by standard red clump stars simultaneously.}


\titlerunning{Non-X-shaped bulge}
\authorrunning{L\'opez-Corredoira et al.}

   \maketitle
%

\section{Introduction}
\label{.intro}

Galactic bulge morphology has been debated over the last decades. The 
first near-infrared surveys showed its axisymmetry \citep[e.g.,][]{Wei94,Lop97}, 
but the shape has not been free from discussion. Its peanut shape in the projected images was interpreted as some imprint of a boxy bulge \citep[e.g.,][]{Dwe95,Lop05}, as predicted by theories that consider stable orbits belonging to several families of periodic orbits \citep[e.g.,][]{Pat03}. However, using  similar images of the Milky Way in the infrared, \citet{Nes16} have more recently claimed to see an X-shaped bulge, 
trying to confirm more recent theories of bulge formation \citep[e.g.,][]{Li15}. 
In spite of their claims, we do not see a clear X-shape in the raw images, and we think 
the processed images may also show elliptical or boxy bulges
depending on the subtraction of some particular disk model \citep{Lop17}, or the subtraction of the bulge as an ellipsoid instead of as a boxy bulge \citep{Han18}.

The structure along the line of sight has also been claimed to show an X-shape using metal-rich red clump giants (RCGs) as standard candles \citep{Nat10,Nat15,McW10,Sai11,Weg13,Sim17}; the structure shows 
a double peak in the star counts for lines of sight within the range of Galactic coordinates of $|\ell|<10^\circ$, $4^\circ<|b|<10^\circ$. However, some doubts have been cast on whether the second peak along the line of sight is a real density structure or an artifact in the luminosity function of red clumps 
\citep{Rat07,Lee15,Lee18,Lee16,Lop16,Lop17,Joo17}. 
There are also signs of a non-X-shaped bulge in other populations: very old and metal-poor stars like RR Lyrae \citep{Pie15}, young ($\lesssim5$ Gyr) populations like F0-F5V stars \citep{Lop16}, Miras variable \citep{Lop17} of all ages (average age equal to 9 Gyr), and metal-poor RCGs \citep{Nes12}. Here, we want to use Gaia-Data Release 2 (DR2) parallaxes in combination with near-infrared 
VISTA (''Visible and Infrared Survey Telescope for Astronomy'') variables in the V\'\i a L\'actea (VISTA-VVV) data
to check the use of RCGs as standard candles, which is the only supporting evidence of an X-shaped bulge.

\section{Data}

\begin{figure}
\vspace{1cm}
\centering
\includegraphics[width=9cm]{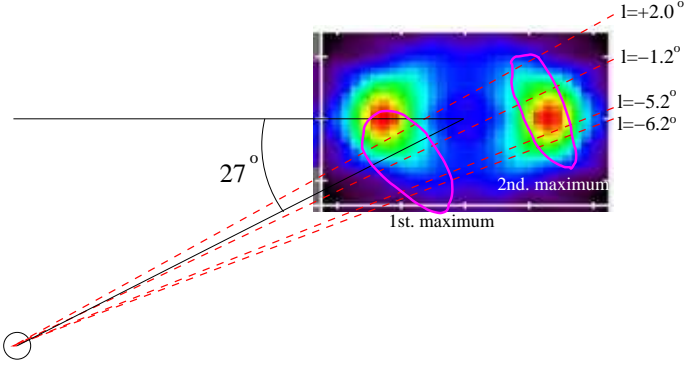}
\caption{Symmetrized density of the X-shaped model at $|z|=1012.5$ pc (equivalent to $|b|=7.3^\circ $
at 8 kpc distance) from \citet[Fig. 12]{Weg13}, 
derived from red clumps, assuming a Galactocentric distance of 8 kpc and a bar angle of 27$^\circ $. 
The dashed lines stand for $\ell =+2.0^\circ, -1.2^\circ, -5.2^\circ, -6.2^\circ $, which cover the range
of our data (see Table \ref{Tab:linessight}). We note how these lines cross two maxima.}
\label{Fig:Wegg2013_fig12}
\end{figure}

VISTA-VVV is an European Southern Observatory (ESO) public survey with the 4.1 m VISTA telescope at Cerro Paranal \citep{Min10,Sai12} in Chile. This telescope performs observations  toward the Galactic bulge in latitudes between -10$^\circ $ and +5$^\circ $ (therefore, we do not have access to the X-shaped features in the positive $b$, and we will only explore the negative ones) and part of the disk. We used the default aperture corrected photometry in 
filters J and Ks. For our analysis, we chose seven lines of sight within the region of interest, those in
\citet{Lop16}, characterized for having the lowest extinction in the areas with Galactic coordinates
$|\ell|\le2^\circ$,
$b=-6.5^\circ$, $-7.5^\circ$, $-8.5^\circ$, $-9.5^\circ,$ and $2^\circ<|\ell|\le 10^\circ $,
 $-7.5^\circ$, $-8.5^\circ, $ $-9.5^\circ$ respectively, according to the
cumulative extinction measurements of \citet{Sch98}. We assume an extinction ratio 
$A_K=0.34\times E(B-V)$ \citep{Sch11}.
Each line of sight covers one square degree
($\cos\delta\Delta\alpha=1^\circ$, $\Delta\delta =1^\circ$)
with its center in the direction indicated in Table \ref{Tab:linessight}; in all of the cases, the extinction
in K-band is very low, lower than 0.1 magnitudes. Models of the X-shape contain the seven lines of sight within the region where the double peak produced by the X-shape should be observed. 
A geometrical description for a particular
model is shown in Fig. \ref{Fig:Wegg2013_fig12}.
Selecting different areas with different values of Galactic coordinates is interesting here in order to show that our measurements are general for the bulge and not placed in a particular line of sight.

The second Gaia data release \citep{Gai18} (Gaia-DR2) consists of astrometry, photometry, parallaxes, proper motions, radial velocities, and information on astrophysical parameters and variability in the full sky. Gaia-DR2 contains celestial positions and the apparent brightness in G magnitude for approximately 1.7 billion sources. For 1.3 billion of those sources, parallaxes and proper motions are in addition available. We selected the sources with magnitude $G\le19$, with completeness above 90\% \citep{Are18}: a total of 0.57 billion sources. We matched the Gaia-DR2 and the VISTA-VVV survey in the selected lines of sight within a maximum separation of 0.20 arcseconds with the algorithm "Sky/topcat": 94.7\% of the Gaia sources have a counterpart in the VISTA-VVV catalog.
The sources with $m_K<14$ are bright enough in VISTA-VVV to have an accurate photometry.

\begin{table}
\caption{Explored lines of sight of an area equal to one square degree.
Columns: 1) Coordinates of the central point of the area; 2) Average extinction; 3) Total number of sources for the
color-magnitude diagram of the cross-correlated survey VVV+Gaia.}
\begin{center}
\begin{tabular}{ccc}
Galactic long., lat. (J2000) & $\langle A_K\rangle$ & $N$ \\ \hline
-5.23$^\circ $, -7.50$^\circ $ & 0.068 & 194\,008 \\
-5.54$^\circ $, -8.50$^\circ $ & 0.047 & 155\,424 \\
-6.16$^\circ $, -9.50$^\circ $ & 0.040 & 104\,856 \\ \hline
2.00$^\circ $, -6.50$^\circ $ & 0.089 & 259\,231 \\
-1.24$^\circ $, -7.50$^\circ $ & 0.078 & 205\,920 \\
-1.24$^\circ $, -8.50$^\circ $ & 0.053 & 172\,818 \\
-1.28$^\circ $, -9.50$^\circ $ & 0.045 & 108\,266 \\ \hline
\label{Tab:linessight}
\end{tabular}
\end{center}
\end{table}

\section{Analysis and results}

\begin{figure}
\vspace{1cm}
\centering
\includegraphics[width=8cm]{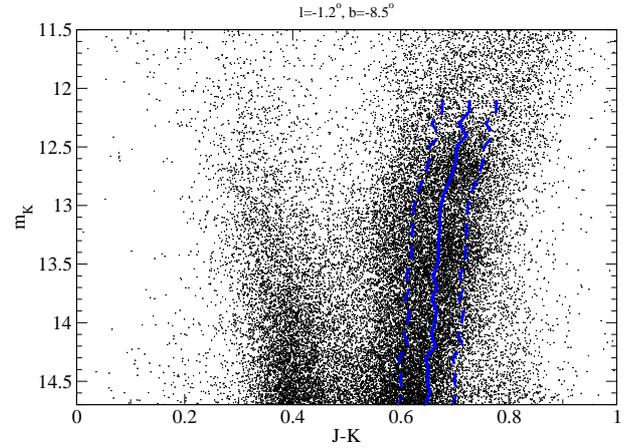}
\caption{Example of color-magnitude diagram for one line of sight at $\ell=-1.2^\circ $, $b=-8.5^\circ $. Between dashed blue lines, we plot the region from which stars were selected. The solid line indicates the average color of stars with $(J-K)>0.55$.}
\label{Fig:CMlm1p2bm8p5}
\end{figure}

\begin{figure}
\vspace{1cm}
\centering
\includegraphics[width=8cm]{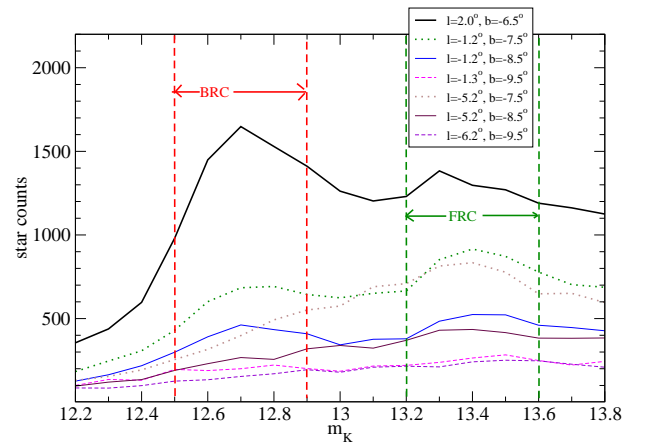}
\caption{Star counts of selected stars in different lines of sight. The peak of bright RCGs is between the two red dashed vertical lines. The peak of faint RCGs is between the two green dashed vertical lines.}
\label{Fig:cb}
\end{figure}

\begin{figure}
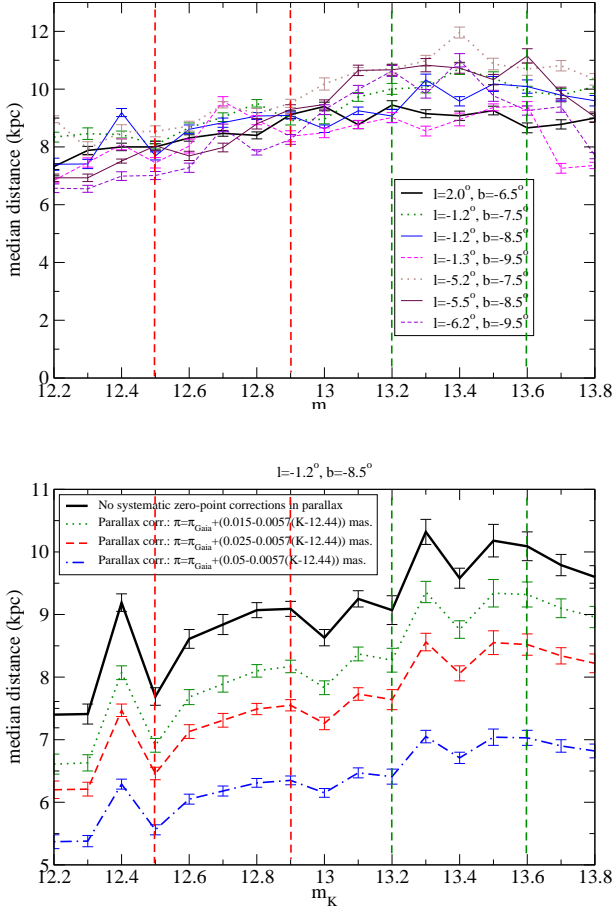

\vspace{1cm}
\centering
\includegraphics[width=8cm]{pb.eps}\\
\vspace{0.5cm}
\includegraphics[width=8cm]{pb2.eps}
\caption{Median distances of RCGs as a function of apparent magnitude. Top panel: Median of distances of the selected stars without
zero-point systematic correction of parallaxes. 
The error bars in that plot represent the error of the average ($\frac{{\rm r.m.s.}}{N-1}$). The peak of bright RCGs is between the two red dashed vertical lines. The peak of faint RCGs is between the two green dashed vertical lines. Bottom panel: Same distance for one line of sight with a zero-point systematic
correction of parallax according to Eq. (\ref{zeropoint}) for $\Delta \pi (G=15)=0.015, 0.025$ and 0.050 mas respectively.}
\label{Fig:pb}
\end{figure}

\begin{figure}
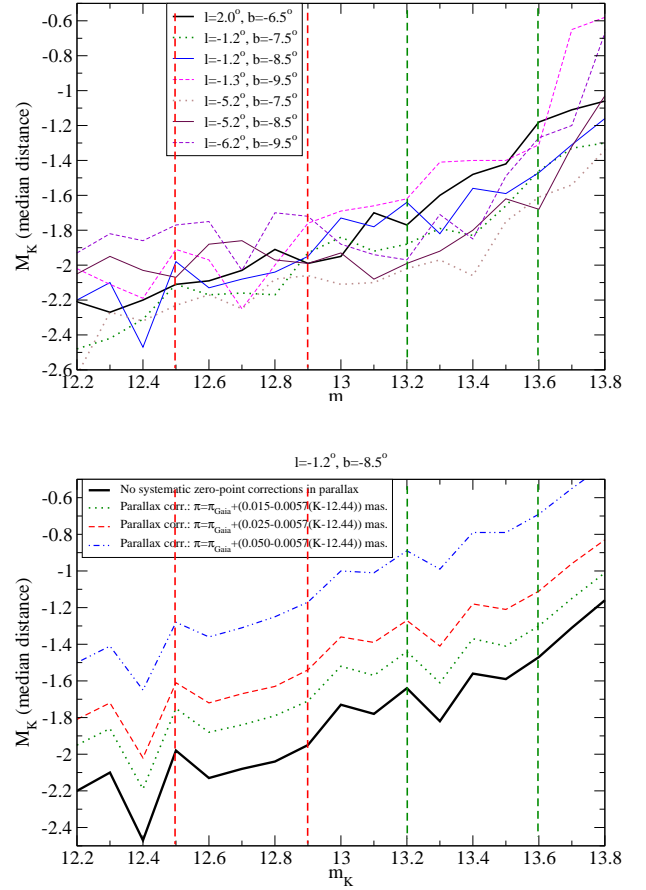

\vspace{1cm}
\centering
\includegraphics[width=8cm]{ab.eps}\\
\vspace{0.5cm}
\includegraphics[width=8cm]{ab2.eps}
\caption{Absolute magnitude corresponding to the median distance of the previous plot. Top panel: Absolute magnitude of the selected stars without zero-point systematic correction of parallaxes. Extinction (from \citet{Sch98}) was taken into account, although it is very low ($<0.1$ mag. in K), so the result will not be affected by the extinction calculation. The peak of bright RCGs is between the two red dashed vertical lines. The peak of faint RCGs is between the two green dashed vertical lines.
Bottom panel: Same absolute magnitude calculation for one line of sight with a zero-point systematic
correction of parallax according to Eq. (\ref{zeropoint}) for $\Delta \pi (G=15)=0.015, 0.025$ and 0.050 mas respectively.}
\label{Fig:ab}
\end{figure}

For each of the seven lines of sight, we derive a K versus J-Ks color-magnitude diagram in near-infrared; an example is given in Fig. \ref{Fig:CMlm1p2bm8p5}. We select the RCGs in this color-magnitude diagrams in the usual way: selecting the right stripe, separated from the main sequence of dwarfs. In particular, we select the stars centered in the line $\langle (J-K)\rangle (m_K)$ for stars with $(J-K)>0.55$ (similar to \citet[Fig. 2]{Sai11}). From this line (solid blue line in Fig. \ref{Fig:CMlm1p2bm8p5}), we take a
range of width $\Delta (J-Ks)=0.1$ to select the RCGs (area between dashed blue lines in 
Fig. \ref{Fig:CMlm1p2bm8p5}).
The use of near infrared is more appropriate than visible for this selection of RCGs, both because of the much lower extinction and because the absolute magnitude of RCGs is almost independent of the age and metallicites of the population beyond a few hundreds of magnitude in K
\citep{Sal02}, so no change of several tenths in absolute magnitude can be attributed to gradients, metallicity, or age within the bulge.
In Fig. \ref{Fig:cb}, we show the star counts of these areas.
The double peak is significantly observed in most cases, although not in all of them. We neglect the contamination of galaxies, whose density is much lower than the density of stars within the present ranges of magnitudes and coordinates.

For each sample of selected stars, we calculate the median of the distance (inverse of the parallax given by Gaia-DR2) as a function of apparent magnitude in K-band (Fig. \ref{Fig:pb}).
The calculation of distance as the inverse of the parallax is not suitable in general for stars at large distances and a Bayesian approach is recommended for it \citep{Lur18,Bai18,Lop19}. However, apart from possible systematic zero-point errors that are treated in Sect. \ref{.syst}, the median distance of a distribution with stars
at the same distance, even with large r.m.s. (but with a very large number of sources), is correctly calculated as the inverse of the median of the parallax with a very small error. Once we select the stars, all assumed to be RCGs with the same parallax $\pi _0$ for a fixed apparent magnitude, the convolution
of a Dirac's delta with a Gaussian in parallax gives an average or median
of the parallax equal to $\pi_0$. Given a constant absolute magnitude and extinction, the distance is fixed. The intrinsic dispersion of absolute magnitudes in RCGs is equal to $0.17\pm 0.02$ mag. \citep{Haw17}, which is negligible.
We use the median instead of the average to characterize the population because it is less affected by the
few outliers of stars with big negative or positive parallaxes \citep{Lur18}. 
All the stars are included for this calculation, even those with negative parallaxes; otherwise the
result would be biased.
The median gives a good representation of the dominant population in the selected sample; ``averages'' are a bit lower than medians but show similar trends. The dispersion of values of the parallaxes is big, with median around 0.04 mas for $m_K=12$ and around 0.08 mas for $m_K=14$; however, since the number of stars per bin is very high, the error in the determination of the average or the median is low.

Figure \ref{Fig:pb} shows some variation of distance with $m_K$, $\frac{\Delta r}{r\Delta m_K}\lesssim 0.2$,
 but this difference is much smaller than 
expected if we were observing populations with constant absolute magnitude for constant $M_K$ (as said,
the intrinsic dispersion of absolute magnitudes in RCGs is negligible)
and $A_K$ ($\frac{\Delta r}{r\Delta m_K}=0.2\,ln(10)$).
The plot of the distance also indicates that the brightest stars ($12.2<m_K<12.8$) come from the closest part
of the bulge at distance 7-9 kpc, whereas fainter stars ($m_K>12.8$) come from the back side of the bulge at 
distances 8-11 kpc. This is what is expected for an ellipsoidal or boxy bulge with a total diameter of 3-4 kpc in
the line of sight, but not for an X-shaped bulge. We will see it more clearly in the following paragraphs.
We also note that differences in the observed radial velocity distribution \citep{Vas13} and Gaia proper motions \citep{San19} between the bright and faint RCGs are explained in a composite bulge model \citep{Lee15,Joo17}, where the pseudobulge in streaming motion is embedded in a classical bulge, without the need of an X-shape. Figure 13 of \citet{San19} shows some differences in Gaia proper motion in the same fields
in which the composite bulge model previously predicted those observed differences \citep{Joo17}. 
It would be interesting to repeat the analysis of radial velocities with Gaia, but that must wait until the release of the Gaia DR3, as the DR2 database only contains velocities for stars up to $G\approx 13$, which is not enough to reach the Galactic bulge.

In Fig. \ref{Fig:ab}, we show the absolute magnitude corresponding to the median distance of Fig. \ref{Fig:pb}; that is, $M_K=m_K-5\log_{10}(r_{\rm median})+5-A_K$. For the extinction, we adopt the total cumulative extinction from \citet{Sch98}; 
we consider negligible the amount of dust at distance $>670$ pc from the plane ($r>6$ kpc) in the bulge region. In any case, the calculated extinctions are very low, $<0.1$ mag. in K, so the result will not be significantly affected by that calculation.  
Clearly, the assumption of $M_K$ equal to -1.6 or -1.7 for all of these stars, 
which has been adopted by all of the groups claiming the discovery of an X-shape in the bulge with RCGs \citep{Nat10,McW10,Sai11,Weg13,Nat15,Sim17}, is not correct. Rather, we must speak of a mixture of populations with different absolute magnitudes.

The faint peak in the counts at Fig. \ref{Fig:cb} has a maximum at $m_K\approx 13.4$, and for this magnitude $M_K\approx-1.63\pm0.04$ (average of all data of $m_K$ between 13.05 and 13.75), so the population of this peak may mostly be composed of normal RCGs because they have precisely this absolute magnitude \citep{Sal02,Haw17}. However, the brighter peak of the counts with maximum at $m_K\approx12.7$ has an associated $M_K\approx-2.02\pm0.03$ (average of all data of $m_K$ between 12.35 and 13.05), which 
certainly cannot correspond to standard RCGs.
Also, for $m_K>13.75$ we have important contamination from other sources different from RCGs, given that the associated $M_K$ is larger than -1.4, but this contamination is well known: it is due to higher ratios of local disk dwarf stars or red giant branch stars with the same colors as the RCGs \citep{Lop02,Lee18}.

This result of a population in the brighter RCGs that is intrinsically $0.39\pm0.05$ magnitudes brighter than the normal RCGs
corroborates the analysis of helium-enhanced RCGs \citep{Lee15}.
The difference of 0.7 apparent magnitudes between the two peaks does not represent the same population of stars that are separated by $\gtrsim 3$ kpc, but different populations that at most are separated by $\sim1.5$ kpc. The construction of the X-shaped bulge density hypothesis cannot be maintained, whereas the hypothesis of the brightest peak composed of
He-enhanced RCGs \citep{Gir99,Lee15,Lee18,Lee16,Joo17} and red giant branch stars \citep{Lee18} agrees excellently with our data.

\section{Systematic errors of parallaxes in Gaia-DR2}
\label{.syst}

Up to now, we have not considered any possible zero-point bias in the parallaxes of Gaia-DR2. However, some 
measurements indicate that there is a systematic bias \citep{Lin18,Are18,Sta18,Zin18,Leu19,Sch19,Hal19} and we should evaluate
its effect on our results.

Let us take for instance
the line of sight $\ell=-1.24^\circ $, $b=-8.50^\circ $:
the median color of the population in the brightest peak at $m_K=12.7$ is $(G-K)=2.6$ and the median color of the population at $m_K=13.4$ is $(G-K)=2.5$. Therefore, the apparent magnitude in Gaia-DR2 G filters is $m_G=15.3$ for the brightest peak and $m_G=15.9$ for the faintest peak. That is, $\Delta m_G=0.6$ between the two peaks.
This difference may change slightly with the line of sight given that it slightly depends on the latitude and the extinction,  but they are small changes in small extinction regions within our area, 
so $\Delta m_G$ is between 0.5 and 0.8 at most and an apparent magnitude of the brightest peak is at $m_G$ between
15 and 15.5.

A systematic error $(\Delta \pi )_{\rm syst}$ in parallax $\pi $ leads to a systematic error in the absolute magnitude equal to
\begin{equation}
\Delta M_{\rm syst}=\frac{5}{ln 10}\frac{(\Delta \pi )_{\rm syst}}{\pi }
,\end{equation}
and the dependence with the magnitude $G$ will be
\begin{equation}
\frac{d\Delta M_{\rm syst}(G)}{dG}=\frac{5}{ln 10}\frac{ \pi (G)\left[\frac{d(\Delta \pi )_{\rm syst}(G)}{dG}\right] - (\Delta \pi )_{\rm syst}(G) \left[\frac{d\pi (G)}{dG}\right]} {\pi ^2(G) }
.\end{equation}

We have between the first and second peak of RCGs
$\left[\frac{d\pi (G)}{dG}\right]\approx -0.03$ mas/mag and $\pi \approx 0.118$ mas in the first peak 
(measured from Fig. \ref{Fig:pb}). Hence,
\begin{equation}
\frac{d\Delta M_{\rm syst}(G)}{dG}\approx 4.7\,(\Delta \pi )_{\rm syst}(G)({\rm mas})
+17.9\,\left[\frac{d(\Delta \pi )_{\rm syst}(G)({\rm mas})}{dG}\right] 
.\end{equation}
The first term gives the systematic variation due to a global systematic zero-point bias in the whole data,
and the second term takes into account the variation with the magnitude.
In the literature, for $G\approx 15$, 
we find values \citep{Lin18} of $(\Delta \pi )_{\rm syst}(G=15)\approx -0.05$ mas and
$\left[\frac{d(\Delta \pi )_{\rm syst}(G=15)({\rm mas})}{dG}\right]\approx +0.0067$ mas/mag (we calculate this amount
as the average derivative between $(\Delta \pi )_{\rm syst}(G=15)\approx -0.05$ and $(\Delta \pi )_{\rm syst}(G=18)\approx -0.03$),
according to Fig. 7, left panel of \citet{Lin18}).
With these values, we get $\frac{d\Delta M_{\rm syst}(G)}{dG}\approx -0.11$. Therefore, the effect
of systematic variation of parallaxes for a difference of $\Delta G=0.6$ would be -0.07 magnitudes, which
is much lower than the $\Delta M_K\approx 0.4$ that we have found between the two red clump peaks.
Color dependence can even reduce $|(\Delta \pi )_{\rm syst}(G=15)|$, since the reddest stars
have a lower value of zero-point systematics \citep[Fig. 7, middle panel]{Lin18}. 
\citet{Hal19} present a specific analysis for RCGs, but with much lower apparent magnitude on average and for many
regions in the sky, so their results cannot be directly applicable for our corrections calculation.
Our variation of absolute magnitude in Fig. \ref{Fig:ab} is observed to be similar in the seven lines of sight with separations up to 
approximately ten degrees, and the spatial variation (rms) at scales of ten degrees is merely $\sim 0.01$ mas \citep[Fig. 14, top panel]{Lin18}, so the spatial fluctuation of the zero point cannot be the explanation either.
Unless much higher values of the systematic errors in parallax and its dependence with the magnitude
arise, we do not think that the full difference of 0.4 magnitudes in absolute magnitude can be
explained by systematics.

The above considerations can be summarized into a systematic correction of the zero point:
\begin{equation}
\label{zeropoint}
\pi=\pi _{\rm Gaia}- [\Delta \pi (G=15)+0.0057(K-12.44)\ {\rm mas}]
,\end{equation}
with $\Delta \pi (G=15)=-0.050$ mas in the worst of the cases, and around -0.025 mas for red stars like our RCGs 
and possible space fluctuations up to -0.015 mas. In Figs. \ref{Fig:pb} (bottom panel) and \ref{Fig:ab} (bottom panel), we illustrate the effect of these corrections in one line of sight for the distance and absolute magnitude, respectively; for other areas, it would be very similar.
As can be observed, the gradient of absolute magnitudes is not removed by these systematic error corrections,
although the calibration of the absolute magnitude and distance would be changed to fainter or closer stars. 
It is of some note that the most reasonable results of the calibration of distance and absolute magnitudes (assuming a median distance around 8 kpc for the bulge and absolute magnitude -1.6 for the faintest peak of RCGs) are for absence or the smallest zero-point corrections.
However, if we assume that the brightest peak of RCGs is the one dominated by the presence of standard RCGs with $M_K=-1.6,$ and the faintest peak is dominated by fainter sources, our data would be compatible with $\Delta \pi (G=15)\sim -0.03$ mas. In any case, as has been stated, our problem of relative difference between the two peaks is almost unaffected. The calculation of the zero point for the particular case of red clumps in the bulge is not our goal here, it is just a by-product of our analysis, which may require further analyses in other future works.

\section{Discussion and conclusions}

Therefore, the conclusion is that an X-shaped bulge hypothesis supported by the RCGs surveys is not valid, 
and it is possible that the structure known as the X-shaped bulge does not exist at all.
There are also signs of a non-X-shaped bulge in other populations (see Sect.  \ref{.intro}).
Using  images of the Milky Way in the infrared, \citet{Nes16}
see an X-shaped bulge in the direction perpendicular to the line of sight. However, as said in the Introduction,
this is possibly an artifact of  subtraction of some particular disk model or the bulge as an ellipsoid instead of as a boxy bulge.
All this evidence together and the lack of trustworthy indicators of an X-shape
indicate that most likely a boxy- or peanut-shaped morphology is a better representation of
the bulge. Nonetheless, the X-shape is not excluded: simply, we have not observed it yet 
and, if it exists, it remains to be discovered.
Possibly, some similar analyses can be applied to other in-plane bulge areas where double peaks of RCGs have also been observed \citep{Nog18,Gon18} and interpreted in some cases in terms of new morphological structures (nothing to do with the X-shape this time), again under the erroneous assumption of RCGs in the bulge being a perfect standard candle.

\begin{acknowledgements}
Thanks are given to the anonymous referee for helpful comments. Thanks are given to Carlos Gonz\'alez-Fern\'andez for exchanging information with us on independent analyses of the parallax distribution red clump giants with VVV+Gaia. Thanks are given to Ruth Chester (language editor of A\&A) for proof-reading of the text. 
MLC and FGL were supported by the grant AYA2015-66506-P of the Spanish Ministry of Economy and Competitiveness (MINECO). YWL acknowledges support from the National Research Foundation of Korea. 
The VVV Survey is supported by the European Southern Observatory, by BASAL Center for Astrophysics and Associated Technologies PFB-06, by FONDAP Center for Astrophysics 15010003, by the Chilean Ministry for the Economy, Development, and Tourism’s Programa Iniciativa Científica Milenio through grant P07-021-F, awarded to The Milky Way Millennium Nucleus. This work has made use of data from the European Space Agency (ESA) mission Gaia (https://www.cosmos.esa.int/gaia), processed by the Gaia Data Processing and Analysis Consortium (DPAC; https://www.cosmos.esa.int/web/gaia/dpac/consortium). Funding for the DPAC has been provided by national institutions, in particular the institutions participating in the Gaia Multilateral Agreement.
\end{acknowledgements}

\end{document}